\documentclass[12pt]{article} 
\usepackage{amsmath,  amsthm, amssymb, latexsym,algorithm,multirow,enumitem,url,booktabs,xcolor,bbm,soul, natbib, fullpage,sectsty, hyperref, graphicx}
\usepackage[title]{appendix}
\newcommand\df{{\text{d}}}   
                             
\newcommand\E{{\text{E}}}           
\newcommand\dg{{\text{dg}}}

\newcommand\mL{{\mathcal{L}}}
\newcommand\mT{{\mathcal{T}}}

\newcommand\N{{\text{N}}} 
 
\newcommand\tr{\text{tr}}   
\newcommand\logit{\text{logit}} 
   
\newcommand{\vc}{\text{vec}}  
\newcommand{\vech}{\text{vech}}

\sectionfont{\large}
\subsectionfont{\normalsize}

\newtheorem{theorem}{Theorem}

\newtheorem{lemma}{Lemma}

\newtheoremstyle{named}{}{}{\itshape}{}{\bfseries}{.}{.5em}{\thmnote{#3}#1}
\theoremstyle{named}
\newtheorem*{namedtheorem}{}

\begin{document}

\title{Second order stochastic gradient update for \\
Cholesky factor in Gaussian variational approximation from Stein's Lemma\footnote{This article is a work in progress and will be updated.}}

\author{Linda S. L. Tan (statsll@nus.edu.sg) \\
Department of Statistics and Data Science\\
National University of Singapore}

\date{}

\maketitle

\begin{abstract}
In stochastic variational inference, use of the reparametrization trick for the multivariate Gaussian gives rise to efficient updates for the mean and Cholesky factor of the covariance matrix, which depend on the first order derivative of the log joint model density. In this article, we show that an alternative unbiased gradient estimate for the Cholesky factor which depends on the second order derivative of the log joint model density can be derived using Stein's Lemma. This leads to a second order stochastic gradient update for the Cholesky factor which is able to improve convergence, as it has variance lower than the first order update (almost negligible) when close to the mode. We also derive second order update for the Cholesky factor of the precision matrix, which is useful when the precision matrix has a sparse structure reflecting conditional independence in the true posterior distribution. Our results can be used to obtain second order natural gradient updates for the Cholesky factor as well, which are more robust compared to updates based on Euclidean gradients.
\end{abstract}

\section{Introduction} \label{sec Introduction}

The Gaussian density is widely used in stochastic variational inference \citep{Hoffman2013} to approximate the true posterior density. In Stan \citep{Stan2019} for instance, an algorithm called automatic differentiation variational inference \citep{Kucukelbir2017} can be used to find a mean-field or full rank Gaussian approximation to the posterior distribution. Use of the reparametrization trick \citep{Kingma2014} yields stochastic gradient updates of the mean and Cholesky factor of the covariance matrix \citep{Titsias2014}, which are able to harness information from the log joint model density directly through its first order derivative. \cite{Tan2018} extended this approach to derive an efficient update for the Cholesky factor of the precision matrix instead, which is useful when the precision matrix is specified to have a sparse structure that mirrors the conditional independence structure present in the true posterior distribution, for hierarchical models such as generalized linear mixed models and state space models. The stochastic gradient updates in the above works depend only on the first order derivative of the log joint model density, which arises naturally when the reparametrization trick is applied.

Stein's Lemma \citep{Stein1981} relates the mean of the function of a random variate that is normally distributed with the mean of its derivative, and it is a powerful tool in statistics with wide applications. Recently, \cite{Lin2019a} demonstrated how Stein's Lemma can be used to derive the first and second order identities in Bonnet's and Price's Theorems respectively \citep{Bonnet1964, Price1958}. They also highlighted the link between Stein's lemma and the reparametrization trick, and used it to derive reparametrizable gradient identities for exponential family mixture distributions such as the skewed Gaussian, Student's $t$ and normal inverse Gaussian. 

In this article, we extend the results of \cite{Lin2019a} and use Stein's Lemma to derive alternative unbiased gradient estimates for the Cholesky factor of the covariance or precision matrix of the multivariate Gaussian, in terms of the second order derivatives of the log joint model density. We show that close to the mode, when the log joint model density can be approximated well by a quadratic function, these second order gradient estimates have variance that is smaller than that of first order updates (almost negligible). Hence second order updates can potentially help to improve convergence in stochastic variational inference.  One disadvantage of using second order gradient estimates for the Cholesky factor is that they require additional storage and computation. However, our initial experiments with logistic regression indicate that this additional effort may be worthwhile due to the greater stability that they bring forth. 

While the above discussion focuses on widely used Euclidean gradients, 
the direction of steepest ascent in the parameter space of a statistical model where distance is measured using the Kullback-Leibler (KL) divergence is actually given by the natural gradient \citep{Amari1998, Amari2016}. The natural gradient is obtained by pre-multiplying the Euclidean gradient by the inverse of the Fisher information matrix and its computation is usually quite complex. However, for a density that is in the minimal exponential family \citep{Wainwright2008}, \cite{Khan2018} showed that the natural gradient of an objective function with respect to the natural parameters can be obtained simply by finding the gradient with respect to the mean parameters, without computing the inverse of the Fisher information matrix explicitly. Combining this approach with results from Bonnet's and Price's Theorems, they derived stochastic natural gradient updates for the mean and precision matrix of the multivariate Gaussian. However, their update of the precision matrix, which is in terms of the second order derivative of the log joint model density, does not ensure that it remains positive definite.

 To overcome this issue, \cite{Tan2022} derived natural gradient updates for the Cholesky factors of the covariance and precision matrices by finding the inverse of the Fisher information matrix analytically.  Natural gradient updates are also provided when the covariance or precision matrix adopt certain sparse structures. Although the natural gradient updates in \cite{Tan2022} are in terms of the first order derivative of the log joint model density due to the use of the reparametrization trick, corresponding second order natural gradients can be obtained easily using the identities that we have derived in this article from Stein's Lemma via a direct substitution.

\section{Notation} \label{notation}
For a $d \times d$ matrix $A$, let $\vc(A)$ denote the $d^2 \times 1$ vector composed of the columns of $A$ stacked in order from left to right. Let $\vech(A)$ denote the $d(d+1)/2$ vector obtained from $\vc(A)$ by eliminating the supradiagonal elements in $A$. Let $L$ denote the $d(d+1)/2 \times d^2$ elimination matrix \citep{Magnus1980} such that $L \vc (A) = \vech(A)$. If $A$ is a lower triangular matrix, then $L^T \vech(A) = \vc(A)$. 

Let $\bar{A}$ denote the lower triangular matrix obtained by setting all elements above the diagonal to zero and $\dg(A)$ the diagonal matrix obtained by setting all non-diagonal elements to zero. We define 
\[
\bar{\bar{A}} = \bar{A} - \dg(A)/2.
\]

Let $\otimes$ denote the Kronecker product. For a $d \times 1$ vector $a$ and a function $f$, we use $\nabla_a f$ to denote $\partial f/\partial a$ and $\nabla_a^2 f$ to denote $\partial^2 f/\partial a \partial a^T$.We also define $e_j$ to be a $d \times 1$ vector with the $j$th element equal to one and zero elsewhere.

\section{Stochastic variational inference} \label{sec SVI}

Let $p(y|\theta)$ denote the likelihood of unknown variables $\theta \in \mathbbm{R}^d$ given observed data $y$, and $p(\theta)$ denote the prior distribution. In variational approximation, the intractable true posterior distribution $p(\theta|y)$ is approximated by a more tractable density $q_\lambda(\theta)$, whose parameter $\lambda$ is chosen to minimize the Kullback-Leibler (KL) divergence between $p(\theta|y)$ and $q_\lambda(\theta)$. Since
\[
\log p(y) = \underbrace{\int q_\lambda(\theta) \log \frac{q_\lambda(\theta) }{p(\theta|y)} \df \theta}_{\text{KL divergence}}  +  \underbrace{\int q_\lambda(\theta) \log \frac{p(\theta, y)}{q_\lambda(\theta)} \df \theta}_{\text{Evidence lower bound}},
\]
minimizing the KL divergence is equivalent to maximizing the evidence lower bound, 
\[
\mL(\lambda) = \E_q[\log p(y, \theta) - \log q_\lambda(\theta)].
\]
When $\mL(\lambda)$ cannot be evaluated in closed form (such as for non-conjugate models), stochastic gradient ascent can be used to maximize $\mL(\lambda)$ with respect to $\lambda$ if an unbiased estimate, $\widehat{\nabla}_\lambda \mL(\lambda)$, of the true gradient $\nabla_\lambda \mL$ is available. Starting from some initial estimate of $\lambda$, an update
\[
\lambda \leftarrow \lambda + \rho_t \widehat{\nabla}\mL(\lambda)
\]
is performed at each iteration $t$. The algorithm will converge to at least a local maximum provided some regularity conditions are met \citep[see][]{Spall2003} and the stepsize satisfies $\sum_{t=1}^\infty \rho_t = \infty$ and $\sum_{t=1}^\infty \rho_t^2 < \infty$.

\subsection{Reparametrization trick}

To obtain an unbiased estimate of $\nabla_\lambda \mL(\lambda)$, note that 
\begin{equation} \label{Eq2}
\nabla_\lambda \mL = \int \nabla_\lambda q_\lambda(\theta)  \{ \log p(y, \theta) - \log q_\lambda(\theta)\} \df \theta -  \int  q_\lambda(\theta)  \nabla_\lambda \log q_\lambda(\theta) \df \theta,
\end{equation}
and the second term is the expectation of the score function, which is zero. Hence 
\begin{equation*}
\nabla_\lambda \mL = \nabla_\lambda \E_q[h(\theta)],
\end{equation*}
where $h(\theta) = \log p(y, \theta) - \log q_\lambda(\theta)$. Note that the gradient operator in the first term of \eqref{Eq2} applies only to $q_\lambda(\theta)$, so $h(\theta)$ can be treated as a function independent of $\lambda$ for the purpose of evaluating $\nabla_\lambda \mL$.

The reparametrization trick \citep{Kingma2014} introduces a differentiable transformation, $\theta = \mT_\lambda(z)$, such that $\lambda$ is contained only within the function $T_\lambda(\cdot)$ while the probability density function $p(z)$ of $z$ is independent of $\lambda$.  Making a variable substitution and applying chain rule,
\begin{equation} \label{Reparametrization trick}
\begin{aligned}
\nabla_\lambda \mL 
&= \nabla_\lambda \int q_\lambda (\theta) h(\theta) \df \theta 
= \nabla_\lambda \int p(z) h(\mT_\lambda(z)) \df z \\
&= \int p(z) \nabla_\lambda \theta \nabla_\theta h(\theta) \df z 
= \E_{p(z)}[ \nabla_\lambda \theta \nabla_\theta h(\theta)].
\end{aligned}
\end{equation}
Hence an unbiased estimate of $\nabla_\lambda \mL$ can be obtained by generating $z$ from $p(z)$ and computing $\widehat{\nabla}_\lambda \mL = \nabla_\lambda \theta \nabla_\theta h(\theta)$. This trick enables gradient information from $h(\theta)$ to be harnessed directly and is useful in reducing the variance of gradient estimates.

\section{First order updates for Gaussian variational approximation} \label{sec_first order updates}

The multivariate Gaussian is a popular choice for approximating the posterior density, and \cite{Titsias2014} derived a stochastic variational algorithm that updates the mean and Cholesky factor of the Gaussian approximation by applying the reparametrization trick. Suppose $q_\lambda(\theta)$ denotes the density function of $\N(\theta|\mu, \Sigma)$. 

\subsection{First order Euclidean gradient updates}
Let $CC^T$ is a Cholesky decomposition of $\Sigma$, where $C$ is a lower triangular matrix. We can consider the transformation $\theta = \mu + Cz$, where $z \sim \N(0, I_d)$ and $\lambda = (\mu^T, \vech(C)^T)^T$. As $\nabla_\mu \theta = I_d$, $\nabla_\mu \mL = \E_{p(z)} [\nabla_\mu \theta \nabla_\theta h(\theta)] =  \E_{p(z)}[\nabla_\theta h(\theta)]$ and a stochastic gradient update of $\mu$ is 
\[
\mu \leftarrow \mu + \rho_t \nabla_\theta h(\theta).
\]
On the other hand, $\nabla_{\vech(C)} \theta = L(z \otimes I)$. Hence
\begin{equation} \label{Euclidean gradient covariance}
\nabla_{\vech(C)} \mL = \E_{p(z)} [L(z \otimes I)\nabla_\theta h(\theta)]
= \E_{p(z)}  \vech(G_1).
\end{equation}
where $G_1 = \nabla_\theta h(\theta) z^T$. This leads to the stochastic gradient update,
\begin{equation*} 
\begin{aligned}
C &\leftarrow C  + \rho_t \bar{G}_1.
\end{aligned}
\end{equation*}

\cite{Tan2018} consider a Cholesky decomposition of the precision matrix $\Sigma^{-1}$ instead in order to exploit the sparsity that is inherently present in the posterior distributions of some hierarchical models. Let $TT^T = \Sigma^{-1}$, where $T$ is a lower triangular matrix. Then $\theta = \mu + T^{-T} z$, where $z \sim \N(0, I_d)$ and $\lambda = (\mu^T, \vech(T)^T)^T$. As $\nabla_{\vech(T)} = - L^T (T^{-1} \otimes T^{-T}z)$, 
\begin{equation} \label{Euclidean gradient precision}
\nabla_{\vech(T)} \mL = - \E_{p(z)} 
[L^T (T^{-1} \otimes T^{-T}z)  \nabla_\theta h(\theta)]
= \E_{p(z)} \vech(G_2).
\end{equation}
where $G_2 = -T^{-T} z\nabla_\theta h(\theta)^T T^{-T}$. This leads to the update,
\begin{equation*} 
\begin{aligned}
T &\leftarrow T + \rho_t \bar{G_2} .
\end{aligned}
\end{equation*}
The updates presented thus far are based on Euclidean gradients. \cite{Tan2022} derived natural gradients for the mean and Cholesky factor of the covariance or precision matrix by finding the inverse of the Fisher information matrix analytically and then pre-multiplying it to the Euclidean gradients obtained from the reparametrization tricks in each case.

\subsection{First order natural gradient updates}
From Theorem 2 of \cite{Tan2022}, the natural gradient with respect to $\mu$ is $\widetilde{\nabla}_\mu \mL = \Sigma \nabla_\theta h(\theta)$, which leads to the natural gradient update,
\[
\mu \rightarrow \mu + \rho_t \Sigma \nabla_\theta h(\theta). 
\]
The difference between this update and the Euclidean gradient update is that $\nabla_\theta h(\theta)$ is pre-multiplied by $\Sigma$ here. 

The natural gradient with respect to $C$, the Cholesky factor of $\Sigma$, is  $\widetilde{\nabla}_{\vech(C)} \mL = \vech(C \bar{\bar{H}}_1)$, where $H_1 = C^T \bar{G}_1$. Note that $\vech(G_1)$ is just the unbiased estimate of the Euclidean gradient of $\mL$ with respect to $\vech(C)$. Hence the natural gradient update for $C$ is 
\begin{equation*}
C \leftarrow C  + \rho_t C \bar{\bar{H}}_1.
\end{equation*}

On the other hand, the natural gradient with respect to $T$, the Cholesky factor of $\Sigma^{-1}$, is  $\widetilde{\nabla}_{\vech(T)} \mL = \vech(T \bar{\bar{H}}_2)$, where $H_2 = T^T \bar{G}_2$ and $\vech(G_2)$ is the unbiased estimate of the Euclidean gradient of $\mL$ with respect to $\vech(T)$. Hence the natural gradient update for $T$ is 
\begin{equation*}
T \leftarrow T  + \rho_t T \bar{\bar{H}}_2.
\end{equation*}

Notably, all the updates above only make use of the first order derivative of $h(\theta)$. Next, we derive alternative updates for the Cholesky factors $C$ and $T$ in terms of the second order derivative of $h(\theta)$ by using Stein's Lemma. We wish to investigate whether the use of second order derivatives will lead to better gradient estimates and improved convergence. Our results extend  Bonnet's Theorem \citep{Bonnet1964} and Price's Theorem \citep{Price1958} involving the mean and covariance matrix of the multivariate Gaussian respectively, to those involving the Cholesky factor of the covariance or precision matrix.

\section{Second order updates for Gaussian variational approximation}
Let $f: \mathbbm{R}^d \rightarrow \mathbbm{R}$ be a function with local absolute continuity as defined in \cite{Lin2019a} and $q_\lambda(\theta)$ denote the density function of $\N(\theta|\mu, \Sigma)$ as before. It suffices to consider $\lambda$ as a vector containing the parameters of $\N(\theta|\mu, \Sigma)$. The precise form of $\lambda$ can be specified depending on the parametrization adopted. \cite{Lin2019a} provide proofs for the first order identity (Bonnet's Theorem) and the second order identity (Price's Theorem) stated below. The second equality in \eqref{Bonnet} is also known as Stein's Lemma. 
\begin{namedtheorem}[Bonnet's Theorem (Stein's Lemma)]
\begin{equation} \label{Bonnet}
\nabla_\mu  \E_q[f(\theta)]  = \E_q[\Sigma^{-1} (\theta-\mu) f(\theta)] =  \E_q[\nabla_\theta f(\theta)],
\end{equation}
\end{namedtheorem}
\begin{namedtheorem}[Price's Theorem]
\begin{equation}\label{Price}
\nabla_\Sigma  \E_q[f(\theta)] = \frac{1}{2} \E_q[\Sigma^{-1}  (\theta-\mu) \nabla_\theta f(\theta)^T] = \frac{1}{2} \E_q [\nabla_\theta^2 f(\theta)],
\end{equation}
\end{namedtheorem}
Note that the second equality in \eqref{Price} can be obtained easily from \eqref{Bonnet} by replacing $f(\theta)$ by $g(\theta) = \nabla_\theta f(\theta)^T e_j$. Then $\nabla_\theta g(\theta) = \nabla_\theta^2 f(\theta) e_j$. Pre-multiplying by $e_i^T$ on both sides, we have 
\[
e_i^T \E_q[\Sigma^{-1} (\theta-\mu) \nabla_\theta f(\theta)^T ] e_j=  e_i^T \E_q[ \nabla_\theta^2 f(\theta)] e_j
\]
for all $1 \leq i, j \leq d$ and thus $\E_q[\Sigma^{-1} (\theta-\mu) \nabla_\theta f(\theta)^T ] = \E_q[ \nabla_\theta^2 f(\theta)]$ since every $(i,j)$ element in these two matrices agree. \cite{Khan2018} derive second order natural gradient updates for $\mu$ and $\Sigma^{-1}$ by first finding the gradient of $\mL$ with respect to the mean of the sufficient statistics corresponding to the natural parameters of the multivariate Gaussian, and then making use of the results in Bonnet's and Price's Theorems. 
 
We extend the theorems of Bonnet and Price to the Cholesky factors of $\Sigma$ and $\Sigma^{-1}$ using Stein's Lemma, and the results are stated in Theorem \ref{thm_Cholesky_identity}. Lemma \ref{lemma1} is instrumental in proving Theorem \ref{thm_Cholesky_identity} and the proofs of Lemma \ref{lemma1} and Theorem \ref{thm_Cholesky_identity} are given in \ref{Appendix A} and \ref{Appendix B} respectively. 

\begin{lemma} \label{lemma1}
\[
\E_q[\{\Sigma^{-1} (\theta-\mu)  (\theta - \mu) ^T  - I_d \}  h(\theta)] =  \E_q[ \nabla_\theta h(\theta)(\theta - \mu) ^T ].
\]
\end{lemma}

\begin{theorem} \label{thm_Cholesky_identity}
Let $CC^T$ and $TT^T$ be the Cholesky decomposition of $\Sigma$ and $\Sigma^{-1}$ respectively, where $C$ and $T$ are lower triangular matrices.
\begin{enumerate}[label={(\alph*)}, leftmargin=1.5em, topsep=3pt, labelsep=3pt, itemsep=1pt]
\item Define $G_1 =  \nabla_\theta h(\theta) (\theta-\mu)^T C^{-T} $ and $F_1 = \nabla_\theta^2 h(\theta) C$. Then $E_q(G_1) = E_q(F_1)$ and
\begin{equation}\label{C identity}
\nabla_{\vech(C)}  \E_q[h(\theta)] = \E_q \vech(G_1) =  \E_q \vech(F_1).
\end{equation}

\item Define $G_2 = - (\theta - \mu) \nabla_\theta h(\theta)^T T^{-T}$ and $F_2 = -\Sigma \nabla_\theta^2 h(\theta) T^{-T}$. Then $E_q(G_2)$ and $E_q(F_2)$ and 
\begin{equation}\label{T identity}
\nabla_{\vech(T)}  \E_q[h(\theta)] = \E_q \vech(G_2) = \E_q   \vech(F_2).
\end{equation}
\end{enumerate} 
\end{theorem}

The results in Theorem \ref{thm_Cholesky_identity} agree with that obtained from the reparametrization trick in \eqref{Euclidean gradient covariance} and \eqref{Euclidean gradient precision}. However, Theorem \ref{thm_Cholesky_identity} also equips us with alternative unbiased estimates of the Euclidean gradients that are based  on the second order derivative of $h(\theta)$. Thus, alternative second order Euclidean gradient updates and natural gradient updates of $C$ and $T$ can be obtained simply by replacing $G_1$ by $F_1$ and $G_2$ by $F_2$ respectively. The stochastic variational inference algorithms thereby obtained are summarized in Tables \ref{Alg1} and \ref{Alg2} respectively.

\begin{table} [htb!]
\centering
\begin{tabular}{|l|l|}
\hline
{\bf Algorithm 1E (Euclidean gradient) }& { \bf Algorithm 1N (Natural gradient)} \\ \hline
\multicolumn{2}{|c|}{
\begin{minipage}{0.8\textwidth}
\vspace{2mm}
Initialize $\mu$ and $C$. For $t=1,2,\dots$,
\begin{enumerate}[leftmargin=1em,topsep=3pt, labelsep=3pt, itemsep=1pt]
\item Generate $z \sim \N(0, I_d)$ and compute $\theta = C z + \mu$.
\end{enumerate}
\vspace{0.5mm}
\end{minipage}
} \\ \hline
\begin{minipage}[t]{0.45\textwidth}
\vspace{-4mm}
\begin{enumerate}[leftmargin=1em,topsep=3pt, labelsep=3pt, itemsep=1pt]
\setcounter{enumi}{2}
\item Update $\mu \leftarrow \mu + \rho_t \nabla_\theta h(\theta)$.
\item 
\begin{enumerate} [leftmargin=1.5em,topsep=3pt, labelsep=3pt, itemsep=1pt]
\item If using first order updates, \\
compute $\bar{G}_1$, where $G_1 = \nabla_\theta h(\theta) z^T$, \\
update $C \leftarrow C + \rho_t \bar{G}_1$.
\item If using second order updates, \\
compute $\bar{F}_1$ where $F_1 = \nabla_\theta^2 h(\theta) C$, \\
update $C \leftarrow C + \rho_t \bar{F}_1$.
\end{enumerate}
\end{enumerate}
 \end{minipage} &
\begin{minipage}[t]{0.45\textwidth}
\vspace{-4mm}
\begin{enumerate}[leftmargin=1em,topsep=3pt, labelsep=3pt, itemsep=1pt]
\setcounter{enumi}{2}
\item Update $\mu \leftarrow \mu + \rho_t  C C^T \nabla_\theta h(\theta)$.
\item 
\begin{enumerate} [leftmargin=1.5em,topsep=3pt, labelsep=3pt, itemsep=1pt]
\item If using first order updates, \\
compute $\bar{G}_1$, where $G_1 = \nabla_\theta h(\theta) z^T$, \\
and $\bar{\bar{H}}_1$ where $H_1 =  C^T \bar{G}_1$.
\item If using second order updates, \\
compute $\bar{F}_1$ where $F_1 = \nabla_\theta^2 h(\theta) C$, \\
and $\bar{\bar{H}}_1$ where $H_1 =  C^T \bar{F}_1$.
\end{enumerate}
\item Update $C \leftarrow C+ \rho_t C \bar{\bar{H}}_1$.
\end{enumerate}
\vspace{-1mm}
 \end{minipage} \\ \hline
\end{tabular}
\caption{Stochastic variational inference algorithms for updating $\mu$ and $C$.}
\label{Alg1} 
\end{table}

\begin{table} [htb!]
\centering
\begin{tabular}{|l|l|}
\hline
{\bf Algorithm 2E (Euclidean gradient)} & {\bf Algorithm 2N (Natural gradient)} \\ \hline
\multicolumn{2}{|c|}{
\begin{minipage}{0.8\textwidth}
\vspace{2mm}
Initialize $\mu$ and $T$. For $t=1,2,\dots$,
\begin{enumerate}[leftmargin=1em,topsep=3pt, labelsep=2pt, itemsep=1pt]
\item Generate $z \sim \N(0, I_d)$ and compute $\theta = T^{-T}z + \mu$.
\end{enumerate}
\vspace{0.5mm}
\end{minipage}
} \\ \hline
\begin{minipage}[t]{0.43\textwidth}
\vspace{-4mm}
\begin{enumerate}[leftmargin=1em,topsep=3pt, labelsep=3pt, itemsep=1pt]
\setcounter{enumi}{2}
\item Update $\mu \leftarrow \mu + \rho_t \nabla_\theta h(\theta)$.
\item 
\begin{enumerate} [leftmargin=1.5em,topsep=3pt, labelsep=3pt, itemsep=1pt]
\item If using first order updates, \\
compute $\bar{G}_2$, where \\
$G_2 = -T^{-T} z \nabla_\theta h(\theta)^T T^{-T}$. \\
Update $T \leftarrow T + \rho_t \bar{G}_2$.
\item If using second order updates, \\
compute $\bar{F}_2$,  where \\
$F_2 = -T^{-T} T^{-1} \nabla_\theta^2 h(\theta) T^{-T}$. \\
Update $T \leftarrow T + \rho_t \bar{F}_2$.
\end{enumerate}
\end{enumerate}
 \end{minipage} &
\begin{minipage}[t]{0.43\textwidth}
\vspace{-4mm}
\begin{enumerate}[leftmargin=1em,topsep=3pt, labelsep=3pt, itemsep=1pt]
\setcounter{enumi}{2}
\item Update $\mu \leftarrow \mu + \rho_t  T^{-T} v$.
\item \begin{enumerate} [leftmargin=1.5em,topsep=3pt, labelsep=3pt, itemsep=1pt]
\item If using first order updates, \\
compute $\bar{G}_2$, where \\
$G_2 = -T^{-T} z \nabla_\theta h(\theta)^T T^{-T}$, \\
and $\bar{\bar{H}}_2$ where $H_2 =  T^T \bar{G}_2$. 
\item If using second order updates, \\
compute $\bar{F}_2$,  where \\
$F_2 = -T^{-T} T^{-1} \nabla_\theta^2 h(\theta) T^{-T}$, \\
and $\bar{\bar{H}}_2$ where $H_2 =  T^T \bar{F}_2$. 
\end{enumerate}
\item Update $T \leftarrow T + \rho_t T \bar{\bar{H}}_2$.
\end{enumerate}
\vspace{1mm}
 \end{minipage} \\ \hline
\end{tabular}
\caption{Stochastic variational inference algorithms for updating $\mu$ and $T$.}
\label{Alg2} 
\end{table}

\subsection{Comparison of the variance of first and second order updates}
Next, we compare the variance of the first order and second order unbiased gradient estimates. Suppose that close to the mode, $\ell(\theta) = \log p(y, \theta)$ can be well approximated by a second order Taylor expansion about its mode $\hat{\theta}$. Then 
\[
\begin{aligned}
h(\theta) &= \log p(y, \theta) - \log q_\lambda(\theta) \\
&\approx \ell(\hat{\theta}) + \frac{1}{2} (\theta - \hat{\theta})^T \nabla_{\theta}^2 \ell(\hat{\theta}) (\theta - \hat{\theta}) + \frac{d}{2}\log(2\pi) + \frac{1}{2}\log|\Sigma| + \frac{1}{2}(\theta - \mu) \Sigma^{-1} (\theta - \mu). \\
\nabla_\theta h(\theta) &\approx \nabla_{\theta}^2 \ell(\hat{\theta}) (\theta - \hat{\theta})  + \Sigma^{-1} (\theta - \mu). \\
\nabla_{\theta}^2 h(\theta) &\approx \nabla_{\theta}^2 \ell(\hat{\theta}) + \Sigma^{-1}. 
\end{aligned}
\]
Thus we have 
\[
\begin{aligned}
G_1 & \approx \nabla_{\theta}^2 \ell(\hat{\theta}) (\theta - \hat{\theta})(\theta-\mu)^T C^{-T}   + \Sigma^{-1} (\theta - \mu)(\theta-\mu)^T C^{-T}, \\
F_1 & \approx \{ \nabla_{\theta}^2 \ell(\hat{\theta}) + \Sigma^{-1} \}C,
\end{aligned}
\]
while
\[
\begin{aligned}
G_2 &= - (\theta - \mu) \{ (\theta - \hat{\theta})^T \nabla_{\theta}^2 \ell(\hat{\theta})  +  (\theta - \mu)^T \Sigma^{-1} \} T^{-T} \\
F_2 &= -\Sigma \{ \nabla_{\theta}^2 \ell(\hat{\theta}) + \Sigma^{-1} \} T^{-T}.
\end{aligned}
\]
Note that the second order updates, $F_1$ and $F_2$, are fixed and hence have zero variance approximately unlike the first order updates which still have positive variance and are subjected to variation arising from the simulation of $z$ from $\N(0, I_d)$.

\section{Application on logistic regression}
In this section, we compare the first and second order gradient updates using a logistic regression model. For the Euclidean gradients, we will use Adam \citep{Kingma2014} to compute the stepsize as the learning rate in Adam is adaptive and can be tailored to each parameter separately. This helps to compensate for the fact that Euclidean gradients are not scaled by the inverse of the Fisher information matrix, unlike natural gradients. In addition, the exponential moving average is used to introduce momentum. For natural gradients, we will also use an alternative approach called stochastic normalized natural gradient ascent with momentum (Snngm) proposed in  \cite{Tan2022}, as it has been observed that Adam does not seem to perform well when paired with natural gradients. As the natural gradient has been scaled by the inverse of the Fisher information matrix, Snngm does not attempt to further perform any  parameter specific adaption unlike Adam. It only helps to adapt the magnitude of the stepsize to the stage of optimization using the norm of the natural gradient, with some momentum added for robustness against noisy gradient. The default setting in \cite{Kingma2014} is used for Adam. We use the same setting for Snngm and the same stopping criterion for terminating the algorithms as in \cite{Tan2022}. All code is written in Julia \cite{Bezanson2017}.

Suppose $y_i|\pi_i \sim \text{Bernoulli}(\pi_i)$ and $\logit(\pi_i) = \theta^T x_i$ for $i=1, \dots, n$. We specify a normal prior $\N(0, \sigma_0^2)$ on $\theta$. Then the log joint model density and its first and second order derivatives are given below.
\[
\begin{aligned}
\log p(y, \theta) &= \sum_{i=1}^n \{y_i \theta^T x_i - \log(1 + \exp(\theta^T x_i)) \} - \frac{d}{2} \log(2\pi \sigma_0^2)- \frac{\theta^T \theta}{2\sigma_0^2}.
\\
\nabla_\theta  \log p(y, \theta) &= \sum_{i=1}^n \left\{ y_i x_i - \frac{\exp(\theta^T x_i)}{1 + \exp(\theta^T x_i)} x_i \right\} - \frac{\theta}{\sigma_0^2}. \\
\nabla_\theta^2  \log p(y, \theta) &= - \sum_{i=1}^n \frac{\exp(\theta^T x_i)}{[1 + \exp(\theta^T x_i)]^2} x_i x_i^T - \frac{1}{\sigma_0^2}I_d. \\
\end{aligned}
\]

\begin{table} [htb!]
\centering
\small
\begin{tabular}{l|c|rrr|rrr|rrr}
\hline
& Algorithm & \multicolumn{3}{c|}{German} & \multicolumn{3}{c|}{Heart} & \multicolumn{3}{c}{ICU} \\ 
 Stepsize &  (Order) & $T$ & $\hat{\mL}$ & time & $T$ &  $\hat{\mL}$ & time & $T$ & $\hat{\mL}$ & time  \\   \hline
\multirow{4}{*}{Adam} & 1E (1) & 14 & -627.5 & 6.7 & 13 & -144.1 & 1.0 & 17 & -115.3 & 1.1 \\
& 1E (2) &  13 & -625.6 & 12.3 & 13 & -144.0 & 1.3 & 16 & -115.2 & 1.2 \\ \cline{2-11}
& 1N (1) & 8 & -625.9 & 5.5 & 9 & -144.1 & 0.8 & 10 & -115.2 & 0.8 \\
& 1N (2) & 8 & -625.6 & 10.2 & 10 & -144.1 & 1.2 & 13 & -115.2 & 1.3 \\ \cline{1-11}
\multirow{2}{*}{Snngm} & 1N (1) & 5 & -625.7 & 3.0 & 6 & -144.0 & 0.4 & 7 & -115.2 & 0.4 \\
& 1N (2) & 4 & -625.6 & 4.9 & 4 & -144.0 & 0.4 & 4 & -115.2 & 0.3 \\ \hline 
\multirow{4}{*}{Adam} & 2E (1) & 44 & -626.0 & 21.1 & 21 & -144.0 & 3.2 & 24 & -115.2 & 3.5 \\
&  2E (2) & 17 & -625.6 & 17.1 & 16 & -144.0 & 3.2 & 21 & -115.2 & 3.7 \\ \cline{2-11}
&  2N (1) & 27 & -625.6 & 22.6 & 22 & -144.0 & 4.9 & 23 & -115.2 & 5.1 \\
&  2N (2) & 15 & -625.6 & 21.2 & 16 & -144.0 & 4.3 & 16 & -115.2 & 4.0 \\  \cline{1-11}
\multirow{2}{*}{Snngm} &  2N (1) & 9 & -625.6 & 6.5 & 10 & -144.0 & 1.8 & 10 & -115.2 & 1.7 \\
& 2N (2) & 4 & -625.6 & 5.0 & 6 & -144.0 & 1.4 & 6 & -115.2 & 1.2 \\ 
\hline
\end{tabular}
\caption{The number of iterations $(T)$ in thousands, estimate of the lower bound at convergence $\hat{\mL}$ and computation time in seconds for each algorithm.} \label{Tab3}
\end{table}

We will consider the three datasets analyzed in \cite{Tan2022}; namely the German credit dataset $(n=1000, d=49)$ and Heart dataset $(n=270, d=19)$ from the UCI Machine Learning Repository and the ICU dataset $(n=200, d=20)$ from \cite{Hosmer2013}. The posterior distribution of $\theta$ is approximated using $N(\mu, \Sigma)$ and a Cholesky decomposition $CC^T$ of $\Sigma$, or $TT^T$ of $\Sigma^{-1}$ is considered. It is actually not meaningful to consider a Cholesky decomposition of the precision matrix here because the computation is more intensive and there is no specific sparsity structure in $\Sigma^{-1}$ that can be be taken advantage of. However, we just wish to investigate and compare the performances of Algorithms 1E, 1N, 2E and 2N. 

Table \ref{Tab3} summarizes the results of the various algorithms, For Algorithm 1E, we note that second order updates led to a higher lower bound in a smaller number of iterations than first order updates, and this is especially so for the German dataset.  However, the overall computation time is increased. For Algorithm 1N, second order updates also took lesser iterations to converge than first order updates when stepsize is computed using Snngm. However, this improvement is absent when Adam is used. For Algorithms 2E and 2N, the advantage of using second order updates is clear across all datasets. Not only is the number of iterations lesser but the overall reduction in computation time is also reduced.

\begin{figure} [htb!]
\centering
\includegraphics[width=0.65\textwidth]{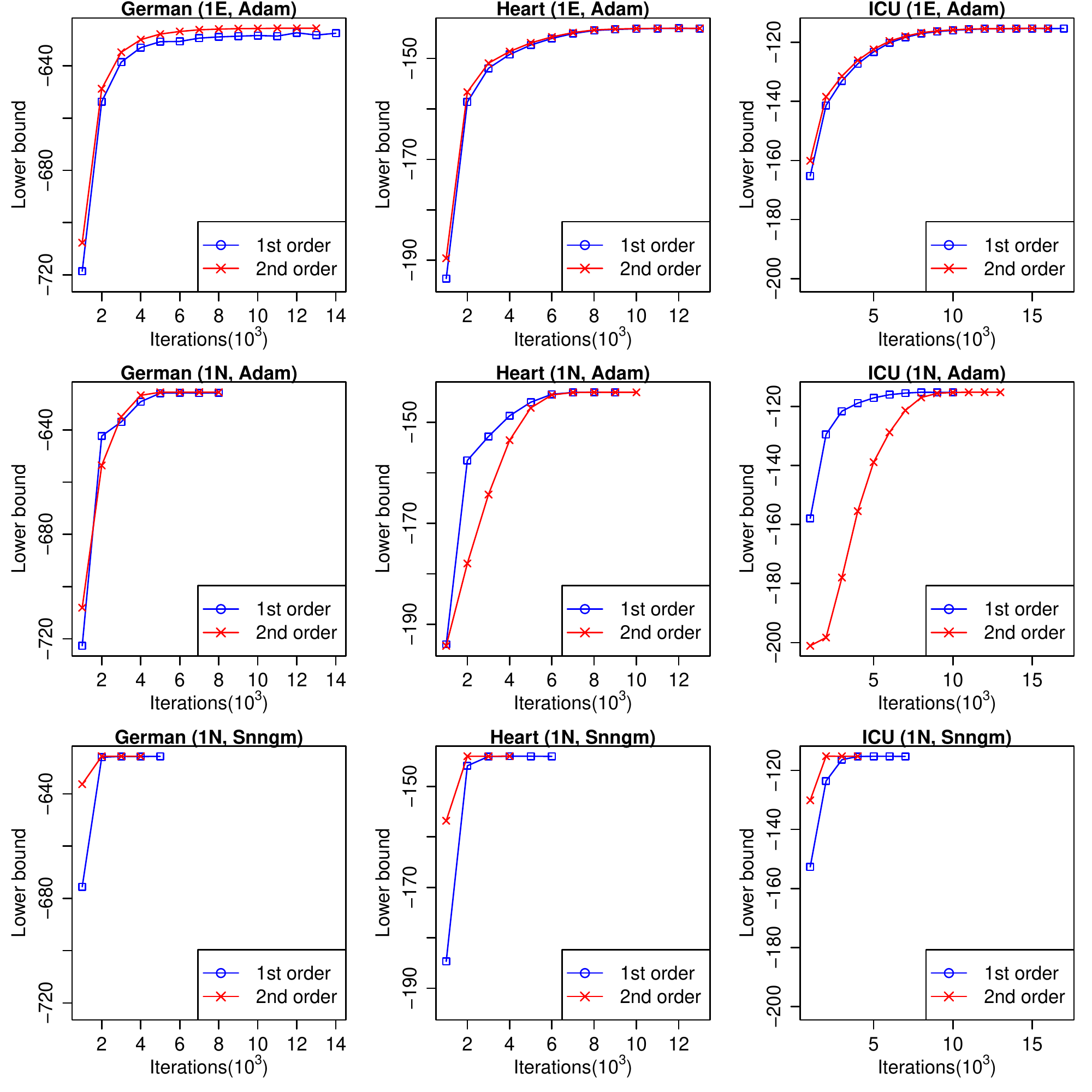}
\caption{Plots show the trajectory of the estimated lower bounds against number of iterations (in thousands) for Algorithms 1E and 1N.} \label{Fig1}
\end{figure}

\begin{figure}[htb!]
\centering
\includegraphics[width=0.7\textwidth]{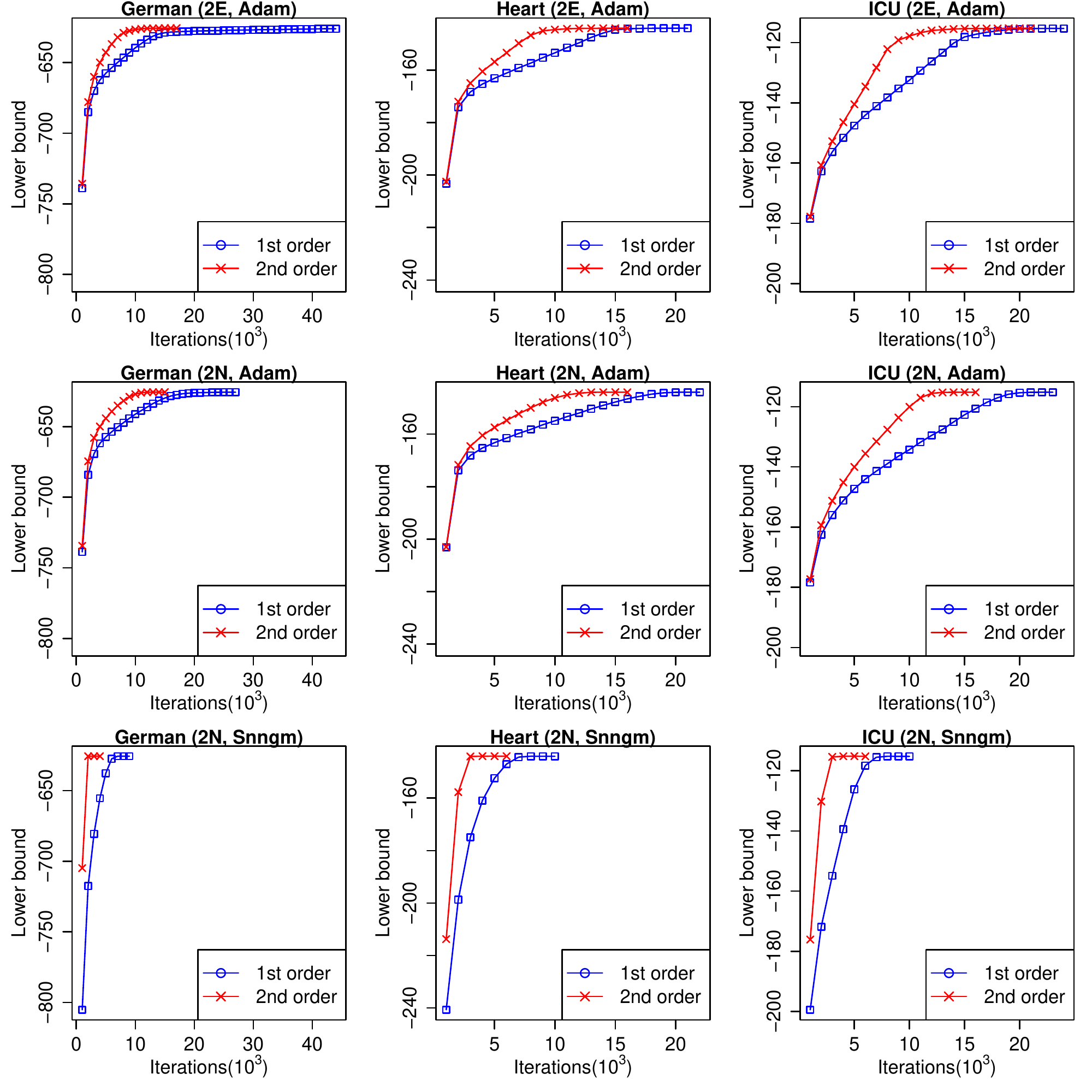}
\caption{Plots show the trajectory of the estimated lower bounds against number of iterations (in thousands) for Algorithms 2E and 2N.} \label{Fig2}
\end{figure}

The trajectories of estimated lower bounds averaged over the past 1000 iterations are shown in Figures \ref{Fig1} and \ref{Fig2}.  The lower bound attained over the first 1000 iteration is generally higher for second order updates than first order ones, except in some cases where the stepsize is computed using Adam. A marked improvement in the trajectory of second order updates can be seen in Algorithm 1N (Snngm) and in Algorithms 2E and 2N.

\section{Conclusion}
In this article, we have introduced alternative second order Euclidean and natural gradient updates for the Cholesky factors of the covariance or precision matrices in Gaussian variational approximation. These second order updates are derived using Stein's Lemma and they are shown to have negligible variance when the log joint model density can be approximated well by a quadratic function close to the mode.  We have investigated the performance of these second order updates in logistic regression across several data sets and they are very useful in improving convergence particularly when a Cholesky decomposition of the precision matrix is considered. We will be updating this article with more results in the future.

\appendix
\renewcommand{\thesection}{Appendix \Alph{section}} 
\renewcommand{\thesubsection}{\Alph{section}.\arabic{subsection}}

\section{Proof of Lemma \ref{lemma1}} \label{Appendix A}
\begin{proof}
Define $g(\theta) = (\theta - \mu) ^T e_j h(\theta)$ to be a function from $\mathbbm{R}^d$ to $\mathbbm{R}$. Then
\[
\nabla_\theta g(\theta) = h(\theta) e_j  +  (\theta - \mu) ^T e_j \nabla_\theta h(\theta).
\] 
Replacing $f(\theta)$ by $g(\theta)$ in Stein's Lemma, we obtain
\[
\E_q[\Sigma^{-1} (\theta-\mu)  (\theta - \mu) ^T e_j h(\theta)] =  \E_q[h(\theta) e_j  +  (\theta - \mu) ^T e_j \nabla_\theta h(\theta)].
\]
This implies that for any $1 \leq i,j \leq d$, 
\begin{equation*}
\begin{aligned}
e_i^T \E_q[\{\Sigma^{-1} (\theta-\mu)  (\theta - \mu) ^T - I_d\}  h(\theta)]e_j =  e_i^T \E_q[ \nabla_\theta h(\theta)(\theta - \mu) ^T] e_j.
\end{aligned}
\end{equation*}
Thus 
\begin{equation*} 
\E_q[\{\Sigma^{-1} (\theta-\mu)  (\theta - \mu) ^T - I_d\}  h(\theta)] =  \E_q[ \nabla_\theta h(\theta)(\theta - \mu) ^T]
\end{equation*}
since every $(i,j)$ element of these two matrices agree with each other.
\end{proof}

\section{(Proof of Theorem 1)} \label{Appendix B}

\begin{proof}
First let $z = C^{-1}(\theta - \mu)$. Then we can write $q(\theta) = (2\pi)^{-d/2}\exp(-z^T z/2)/|C|$. Differentiating $q(\theta)$ with respect to $C$ using vector differential calculus \citep{Magnus2019}, 
\begin{equation*}
\begin{aligned}
\df q(\theta) &= q(\theta) [ - \tr(C^{-1} \df C) + z^T C^{-1} (\df C) z] \\
&= q(\theta) \vc (C^{-T} zz^T -  C^{-T})^T L^T \df \vech(C). \\
\therefore \nabla_{\vech(C)} q(\theta)  & = q(\theta) \vech \{\Sigma^{-1}(\theta -\mu)(\theta-\mu)^TC^{-T} -  C^{-T}\}.
\end{aligned}
\end{equation*}
Hence
\begin{equation} \label{Eq1}
\begin{aligned}
\nabla_{\vech(C)}  \E_q[h(\theta)] &= \int \nabla_{\vech(C)} q(\theta)  h(\theta) \df \theta\\
& = \E_q[ \vech \{\Sigma^{-1}(\theta -\mu)(\theta-\mu)^TC^{-T} -  C^{-T}\} h(\theta)].
\end{aligned}
\end{equation}
Post-multiplying the identity in Lemma \ref{lemma1} by $C^{-T}$, we obtain
\[
\E_q[\{\Sigma^{-1} (\theta-\mu)  (\theta - \mu) ^T C^{-T} - C^{-T}\}  h(\theta)] =  \E_q[ \nabla_\theta h(\theta)(\theta - \mu) ^T C^{-T}]
\]
Hence, we have  
\[
\nabla_{\vech(C)}  \E_q[h(\theta)] = \vech\{\E_q[ \nabla_\theta h(\theta)(\theta - \mu) ^T C^{-T}]\}
\]
from \eqref{Eq1}, and the first part of the identity in \eqref{C identity} is shown. For the second part of the identity, we have $\E_q[\Sigma^{-1} (\theta-\mu) \nabla_\theta h(\theta)^T]= \E_q[\nabla_\theta^2 h(\theta) ]$ from Price's Theorem in \eqref{Price}. Taking the transpose and post-multiplying by $C$, we obtain the second part of the identity in \eqref{C identity},
\[
\E_q[ \nabla_\theta h(\theta)  (\theta-\mu)^T C^{-T}]= \E_q[\nabla_\theta^2 h(\theta)C ].
\]

Next, let $z = T^T (\theta - \mu)$. Then we can write $q(\theta) = (2\pi)^{-d/2}|T|\exp(-z^T z/2)$. Differentiating $q(\theta)$ with respect to $T$, 
\begin{equation*}
\begin{aligned}
\df q(\theta) &= q(\theta) [ \tr(T^{-1} \df T) - z^T (\df T)^T (\theta - \mu)] \\
&= q(\theta) \vc (T^{-T} -  (\theta-\mu) z^T)^T L^T \df \vech(T). \\
\therefore \nabla_{\vech(T)} q(\theta)  & = q(\theta) \vech \{T^{-T} -  (\theta-\mu)(\theta - \mu)^T T\}.
\end{aligned}
\end{equation*}
Hence
\begin{equation} \label{Eq4}
\begin{aligned}
\nabla_{\vech(T)}  \E_q[h(\theta)] &= \int \nabla_{\vech(T)} q(\theta)  h(\theta) \df \theta\\
& = \E_q[ \vech \{T^{-T} -  (\theta-\mu)(\theta - \mu)^T T\} h(\theta)].
\end{aligned}
\end{equation}
Taking the transpose of the identity in Lemma \ref{lemma1} and post-multiplying by $T^{-T}$,
\[
\E_q[\{ (\theta-\mu)  (\theta - \mu) ^T T - T^{-T}\}  h(\theta)] =  \E_q[ (\theta - \mu) \nabla_\theta h(\theta)^T T^{-T}].
\] 
From \eqref{Eq4}, this implies that 
\[
\nabla_{\vech(T)}  \E_q[h(\theta)] = - \vech \{\E_q[ (\theta - \mu) \nabla_\theta h(\theta)^T T^{-T}]\}
\]
and the first part of the identity in \eqref{T identity} is shown. For the second part of the identity, we have $\E_q[\Sigma^{-1} (\theta-\mu) \nabla_\theta h(\theta)^T]= \E_q[\nabla_\theta^2 h(\theta) ]$ from Price's Theorem (\eqref{Price}). Pre-multiplying by $\Sigma$ and post-multiplying by $T^{-T}$, we obtain the second part of the identity in \eqref{T identity},
\[
\E_q[(\theta-\mu) \nabla_\theta h(\theta)^TT^{-T}]= \E_q[\Sigma \nabla_\theta^2 h(\theta) T^{-T}].
\]
\end{proof}


\vskip 0.2in
\bibliographystyle{chicago}
\bibliography{refnatgrad}

\end{document}